\documentclass{aircc}

\usepackage{mathpartir}
\usepackage{hyperref}
\usepackage{listings}

\usepackage{booktabs}
\usepackage{cite}
\usepackage{amsmath,amssymb,amsfonts}
\usepackage{graphicx}
\usepackage{textcomp}
\usepackage{xcolor}
\usepackage{multirow}

\usepackage{comment}

\usepackage{circuitikz}

\usepackage{caption}
\usepackage{subcaption}

\usetikzlibrary{shapes,positioning}
\usepackage{tikz}
\usepackage{multicol}
\usepackage{lipsum,adjustbox}

\usepackage{multirow, makecell}

\usepackage{diagbox}

\usepackage{algorithm} 
\usepackage{algpseudocode}

\usepackage{pgf-umlsd}

\title{Verifying Outsourced Computation in an Edge Computing Marketplace}

\author{Christopher Harth-Kitzerow$^*$, Gonzalo Munilla Garrido}
\affiliation{Technical University of Munich  \\
	\email{$^*$\href{mailto:christopher.harth-kitzerow@tum.de}{christopher.harth-kitzerow@tum.de}}
}

\begin{document}


\maketitle

\begin{abstract}
An edge computing marketplace could enable IoT devices (Outsourcers) to outsource computation to any participating node (Contractors) in their proximity. In return, these nodes receive a reward for providing computation resources. In this work, we propose a scheme that verifies the integrity of arbitrary deterministic functions and is resistant to both dishonest Outsourcers and Contractors who try to maximize their expected payoff. We tested our verification scheme with state-of-the-art pre-trained Convolutional Neural Network models designed for object detection. On all devices, our verification scheme causes less than 1ms computational overhead and a negligible network bandwidth overhead of at most 84 bytes per frame. Our implementation can also perform our verification scheme's tasks parallel to the object detection to eliminate any latency overhead. Compared to other proposed verification schemes, our scheme resists a comprehensive set of protocol violations without sacrificing performance.
\end{abstract}




\section{Introduction}\label{chapter:introduction}

Offloading computational tasks from IoT devices to computation resources at the Edge can improve the responsiveness of existing applications and enable the implementation of novel latency-sensitive use cases \cite{offloading_method_P2p}.  
In an edge computing marketplace, we assume that the Outsourcer is a computationally weak IoT device that outsources real-time data to a Contractor to process. The Contractor can be an edge server or any device in proximity to the Outsourcer with enough computational resources available to execute the assigned function reliably and with low latency.

Compared to fixed client-server assignments, an Edge computing marketplace has the potential to overcome the challenges of limited availability of servers, insufficient quality of service, and idle server resources over longer periods of times. Dynamic assignments in an open marketplace can lead to increased availability and competition among edge servers. This allows IoT applications to profit from increased connectivity and responsiveness due to better matching. Edge servers, on the other hand, profit from higher resource utilization due to increased matching rates. 

As IoT devices are often computationally weak, it might be difficult for them to verify whether responses returned by a third-party Contractor are valid. In fact, Contractors have an incentive to return a computationally less expensive probabilistic result to save resources while still collecting the reward. Likewise, an Outsourcer has no incentive to pay an honest Contractor right after receiving all computational results, and expensive micro-transactions prohibit real-time payment.

In this work, we introduce a verification scheme for computation marketplaces that can verify the integrity of arbitrary functions. The Outsourcer verifies the integrity of returned responses by sending some inputs to another Contractor in proximity called the Verifier. We refer to this approach as sampling-based re-execution. We evaluate our scheme's performance with outsourced object detection based on a real-time image stream sent by an IoT device to an edge server. We provide the following contributions:
\begin{enumerate}
    \item We compile a comprehensive list of potential threats that this re-execution based approaches might be vulnerable to in the presence of dishonest participants.
    \item We address all identified threats, by combining existing techniques proposed by related work as well as introducing two novel ones.
    \item Our resulting verification scheme requires little TTP interaction and is resistant to dishonest Outsourcers, Contractors, and Verifiers. The TTP does not have to be located at the edge and can act with arbitrary latency.
    \item Our implementation demonstrates that our verification scheme causes negligible communication and latency overhead. 
\end{enumerate}

\section{Design of our Verification Scheme}\label{chapter:designed_verification_scheme}

As our scheme uses re-execution as a verification approach, it can be applied to any deterministic function. We assume the following setting for outsourced object detection in an edge computing marketplace.
\begin{enumerate}
    \item Edge servers are stationary (reappearing actors) and offer outsourced computation for a fee. They can either act as Contractors or Verifiers.  
    \item Outsourcers are mobile (reappearing and adhoc actors).
    \item Outsourcers owe a reward to Contractors and Verifiers for each processed input.
    \item Each edge participant may act dishonestly but tries to maximize its expected payoff.
    \item A TTP or Blockchain is present that provides a public key infrastructure, a reputation system, and handles payments. This payment settlement entity does not have to be located at the edge. 
    
\end{enumerate}

In the initial situation, we assume an Outsourcer and a Contractor have agreed to a contract. The contract contains a unique ID that distinguishes it from previous contracts with the same participant and contains the participants' public keys registered at the payment settlement entity. Also, a reward per processed input and the function/model to be used is specified. Additionally, the Contractor and Outsourcer may agree on fines, deposits, and bounties if a participant is caught cheating to increase the protocol's robustness. Also, we assume multiple Verifiers are available and willing to agree on a contract with the Outsourcer to process random sample inputs.

\subsection{Preparation Phase}

The preparation phase is responsible for assigning a Verifier to a contract while preventing collusion. We describe planned collusion as the scenario where two participants, either a Contractor and a Verifier, or a Contractor and an Outsourcer, know each other initially and try to trick the other participant by colluding. We describe ad-hoc collusion as the scenario where a Contractor and a Verifier do not initially know each other but still try to communicate and collude.

\subsubsection{Randomization}

\begin{figure}
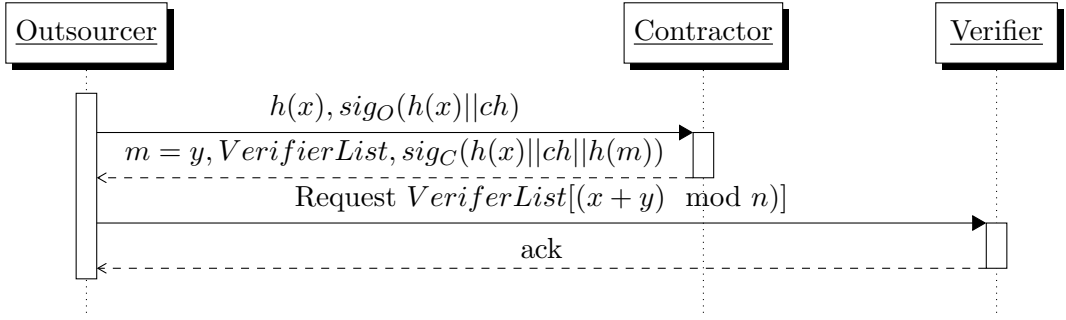

    \centering

\begin{sequencediagram}
        \newthread[white]{o}{Outsourcer}
        \newinst[6]{c}{Contractor}
        \newinst[2]{v}{Verifier}

            \begin{call}
            {o}{$h(x), sig_O(h(x)||ch)$}
            {c}{$m = y, VerifierList , sig_C(h(x)||ch||h(m))$}
            \end {call}
            
                        \begin{call}
            {o}{Request $Verifer List[(x+y)\mod n)]$ }
            {v}{ack}
           
            \end {call}

    \end{sequencediagram}
    
        \caption{Randomization}
    \label{fig:randomization}
    
    \end{figure}

Randomization ensures that the Outsourcer and the Contractor commit to a random Verifier. The Verifier and the Contractor do not learn each other's identity. The protocol consists of the following steps: The Outsourcer digitally signs the hash $h(x)$ of a large random number $x$ and the contract hash $ch$. It sends $h(x)$ and the signature to the Contractor without revealing $x$ itself. The Contractor digitally signs the received hash of the Outsourcer along with a large random number $y$, the contract hash, and a list of available Verifiers sorted by their public keys. Along with this signed hash, the Contractor sends the value $y$ and the list of available Verifiers to the Outsourcer. By signing the initially sent hash of the Outsourcer, the Contractor commits to $x$ and $y$ without knowing $x$. Figure \ref{fig:randomization} illustrates this protocol.

If the list of available Verifier matches the current local list of the Outsourcer to a certain degree, and the signature matches with all revealed messages, the Outsourcer contacts the Verifier at $(x+y)\mod n$ in the list, where $n$ is the total number of Verifiers. If the Outsourcer contacts a different Verifer it will not be able to present the necessary Contractor signatures during Contestation.

\subsubsection{Game-theoretic Incentives}

Even if the Verifier and the Contractor do not know each other, there is a risk of ad-hoc collusion. This is the case, for example, if there exists a $q$-algorithm that is computationally inexpensive but provides a correct result with a certain probability $q$. For object detection, this might be the naive response that no object was found and, therefore, no bounding boxes at specific coordinates have to be estimated and returned. Suppose there is a reward $r$ for returning a valid result and computation costs when computing the desired function honestly $c_h$ or dishonestly $c_d$. From a game-theoretic perspective \cite{game-theory-introduction}, there are two nash equilibria \cite{robust-game-theory}. One nash equilibrium exists when both players act honestly, but the other when both players act dishonestly \cite{belenkiy2008incentivizing} \cite{kupcu}. 

It is crucial to design incentives that eliminate the Nash equilibrium of both players acting dishonestly. Verification schemes such as \cite{kupcu} and \cite{belenkiy2008incentivizing} have identified a relationship of incentives by adding fees for cheating players and bounties for dishonest players such that being honest is a dominant strategy from a game-theoretic point of view. Figure \ref{tab:Contractor_advanced} illustrates the payoff matrix of the Contractor with the use of a bounty $b$ and a fee $f$. The payoff matrix for the Verifier looks identical. Our scheme uses an initial deposit to enforce that a cheating participant pays its fine after being detected. However, the participants have the flexibility to agree on a contract that differs from the ideal incentives discussed here.

\begin{table}[b]
    \centering
    \caption{Payoff matrix with honesty-promoting incentives}
    \begin{tabular}{|l|c|c|}
    \hline
        \backslashbox{Contractor}{Verifier} & Diligent & Dishonest \\ \hline
        \multicolumn{1}{|l|}{Diligent}  & $r-c_h$ & $r-c_h + b$ \\  \hline
        \multicolumn{1}{|l|}{Dishonest} & $rq - (f+b)(1-q) - c_d$ & $r - c_d$ \\ \hline
    \end{tabular}
    
    \label{tab:Contractor_advanced}
\end{table}

\subsection{Execution Phase}

After agreeing on a contract with the random Verifier, the Outsourcer starts sending input to the Contractor and the Verifier to process.  

\subsubsection{Sampling}

Sampling refers to picking one random input out of a collection of inputs. In our verification scheme, the Outsourcer can send samples to the Verifier to check whether its response matches the Contractor's response belonging to the same input. We call this process sampling-based re-execution. Sampling-based re-execution has a significant advantage over complete re-execution. Just with a few samples, a dishonest Contractor with a cheating rate of $c$ can be detected with nearly 100\% confidence. Thus, we can significantly improve the efficiency of the verification process at a negligible security drawdown.

During sampling, the Outsourcer splits up the number of inputs in $i$ intervals and sends only one random sample per interval to the Verifier. In this case the chance $p$ of detecting a cheating attempt is $p = 1 - (1-c)^i$. Even if the Contractor has a low cheating rate, e.g., $c = 0.1$, the computational overhead of detecting cheating with 99\% confidence is less than 0.5\%. While this resulting computational overhead is already low, further optimization could include lowering the sampling rate the longer the contract goes on, or adjusting the sampling rate when a global reputation system is present.

\subsubsection{Digital Signatures}

Since participants communicate with unmonitored peer-to-peer communication in our verification scheme, we need a way to securely record payment promises and dishonest behavior. Otherwise, an Outsourcer could claim never to have received any responses from the other participants. Likewise, Contractor and Verifier could deny that a dishonest result originated from them and could claim to have processed more responses than they did. A payment settlement entity cannot solve a dispute and hold entities accountable without tamper-proof records. 

When an Outsourcer sends an input, it always attaches a digital signature signed over the current input index, the contract hash, and the input itself. This ensures that each signature can be traced back to one unique input in a contract context. Also, the Outsourcer includes a number of currently acknowledged outputs $n$ to the message and signature. Thus, the Contractor and the Verifier receive a signed commitment of redeeming $n$ times the specified reward per response. The unique contract hash ensures that each participant can only redeem payment once per contract.  

\begin{figure}[ht]
	\centering
  \includegraphics[width=0.65\textwidth]{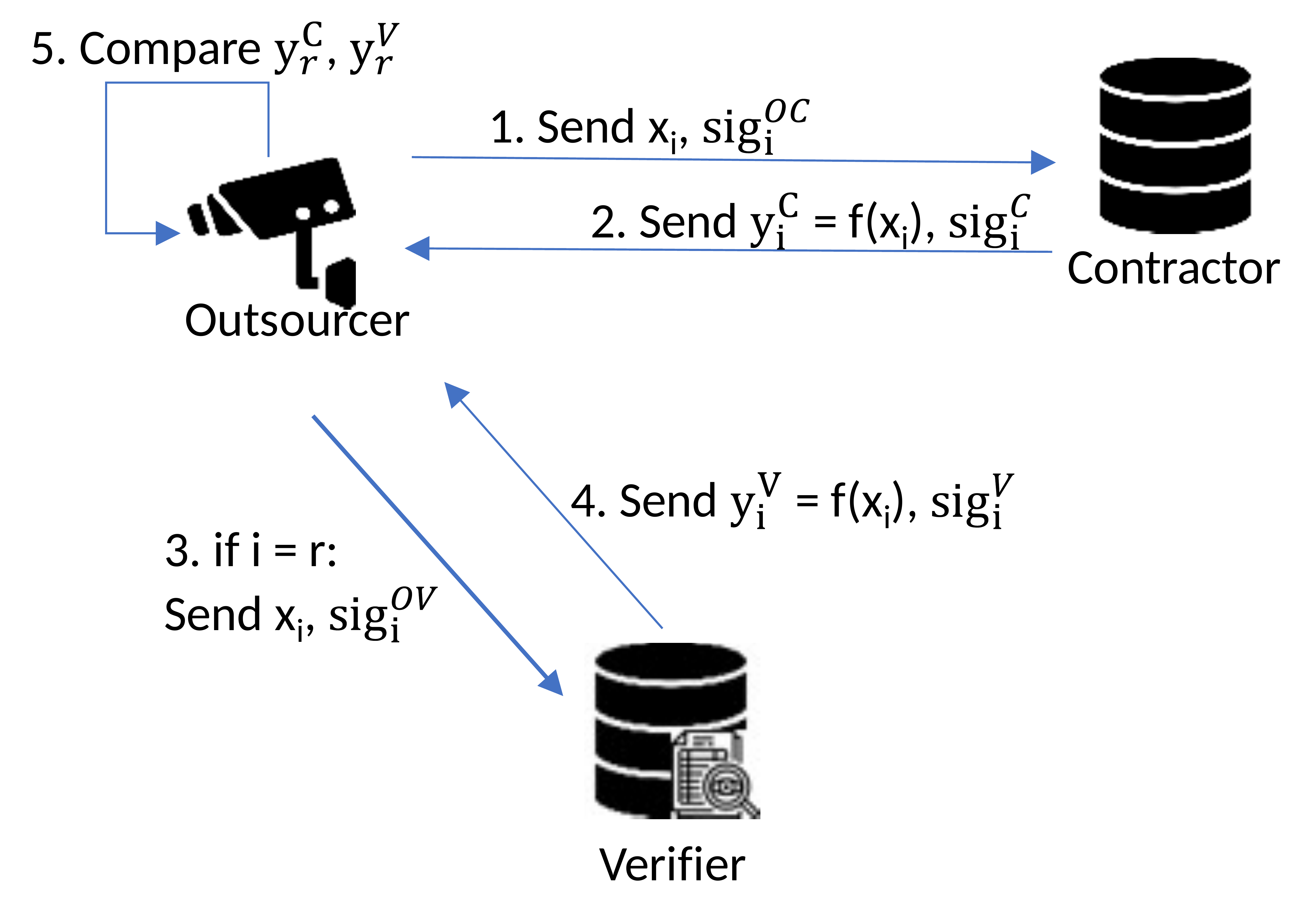}
	\caption{Execution phase}
	\label{signatures}
\end{figure}

When the Contractor or the Verifier send a result to the Outsourcer, they attach the associated input index, along with a digital signature forged over contract hash, input index, and the input itself. This signature serves as proof of being the originator of a false message when detected cheating. If the signature verification fails, a participant can abort the contract according to a QoS violation. It may also leave a bad review to the participant and blacklist it. Figure \ref{signatures} shows a high-level overview of sampling in combination with digital signatures. "$i=r$" refers to the property of our scheme that an input $x$ is only sent to the Verifier if its index $i$ is chosen as a random number $r$ within the current interval. 


\subsubsection{Commitment to Messages Using Merkle Trees}
Merkle Trees can be used as a data structure to efficiently verify whether data is contained in a large collection. The root hash of a Merkle tree can be used to verify if data is included in the whole tree with $log_2(n)$ steps. Verification schemes such as \cite{nabi} use this attribute of Merkle trees in combination with signatures or TTPs to commit to large amounts of inputs or outputs by sending the signed Merkle tree root hash. 

Committing to responses via a Merkle Tree has the advantage that the Outsourcer can commit to a collection of responses before receiving a proof-of-membership challenge by the Outsourcer. When receiving a challenge from the Outsourcer, the Contractor cannot quickly re-compute a response and send it instead since it would not meet the proof of membership challenge. This way, instead of signing all outputs, it is sufficient for the Contractor only to sign Merkle root hashes and challenges in a specified interval, thus improving efficiency.

\subsection{Closing Phase}

A participant can terminate a contract at any time if it sends a termination message to each counter-party. If the contract is terminated according to custom, the Verifier and the Contractor store their last signed input of the Outsourcer, containing the latest number of acknowledged outputs and the signed contract hash. They send the whole signed input, each value over which the signature was forged, and the contract to the payment settlement entity. To prevent Quality of Service (QoS) violations, a global reputation system and local blocklists per node ensure that participants that provide reliable service can be identified. Thus, all participants can submit a review for the other participants at the end of a contract. 

The payment settlement entity simply has to check whether all inputs can be verified by the sent signature. Then, it deducts the reward per input specified in the contract times the number of acknowledged output contained in the last input on behalf of the Outsourcer after a deadline. Within that deadline, the Outsourcer can abort a contract due to dishonest behavior in case a response sent by the Verifier and a response sent by the Contractor belonging to the same input are unequal. In this case, the Outsourcer sends the input, both responses, their signatures, and the contracts to the payment settlement entity. For scalability reasons, the payment settlement entity does not re-execute the input. It only checks if all values match their signatures and verifies if responses are indeed unequal. Provisionally, the Contractor is accused of cheating. Within the specified deadline, the Contractor can decide to engage in a protocol we call Contestation to prove that the Verifier's response was incorrect instead.

\subsubsection{Contestation}

We designed the Contestation protocol to ensure that an honest Contractor or Verifier that is falsely accused of cheating can prove its innocence. A participant accused of cheating may decide to re-outsource the original input to two additional random Verifiers within a deadline. If both random Verifiers return a response that matches the participant's response, it presents their responses and signatures to the payment settlement entity. The participant out of the Contractor and the Verifier that has the minority of random Verifier support is then accused of cheating instead. 

If a Verifier is accused of cheating, it can use the identical protocol to contact two additional random Verifiers and flip the majority of random Verifier support. This protocol might be repeated until no available Verifiers are left. In that case, the participant having the majority of Verifier responses matching with theirs is assumed to be honest. The other participant is assumed to be dishonest and gets fined. If more than 50\% of available Verifiers are non-colluding, Contestation serves as a guarantee that the participant who responded with a false response is found guilty. In combination with a nearly 100\% detection rate of cheating using sampling-based re-execution, any cheating parting will be eventually found guilty with high probability. The combination of Contestation and sampling is designed to make any cheating attempt irrational. The participant that is found guilty at the end of the protocol has to not only pay the specified fee in its contract but also the additionally consulted Verifiers. In the first round of Contestation, the Outsourcer first has to proof to the payment settlement entity that it contacted the correct Verifier by presenting the received Contractor signatures.  

\begin{figure}
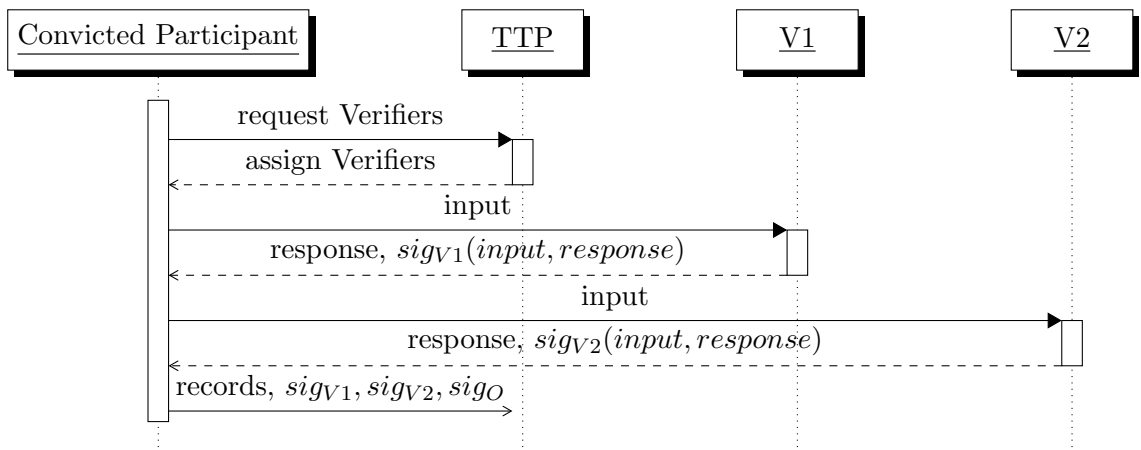

    \centering

\begin{sequencediagram}
        \newthread[white]{p}{Convicted Participant}
        \newinst[2]{t}{TTP}
        \newinst[2]{v1}{V1}
        \newinst[2]{v2}{V2}

            \begin{call}
            {p}{request Verifiers}
            {t}{assign Verifiers}
            \end {call}
            
            \begin{call}
            {p}{input}
            {v1}{response, $sig_{V1}(input,response)$}
            \end {call}
            
            \begin{call}
            {p}{input}
            {v2}{response, $sig_{V2}(input,response)$}
            \end {call}

            \mess{p}{records, $sig_{V1}, sig_{V2},  sig_{O}$ }{t}

    \end{sequencediagram}
    
        \caption{Contestation}
    \label{fig:Contestation}
    
    \end{figure}

Figure \ref{fig:Contestation} illustrates a message sequence chart of Contestation. Notice that the TTP is involved in a minimal amount of computation to ensure scalability. Only if the convicted particpant submits the last message it needs to verify whether the records the contractor provides match the Outsourcer's signature and whether Verifier signatures were formed over the same input value as the original message. In cases where one of the convicted participant does not get a majority of Verifier support it has no incentive to send the last message and occupy the TTP.  

Note that only for an innocent participant it is rational to perform Contestation as additional random Verifiers have to be paid for their service. Assuming an honest ecosystem, a cheating participant that is performing Contestation is wasting additional monetary resources. The computational overhead of Contestation is low as only one input has to be recomputed. However, it requires finding multiple available Verifiers in the system. As latency is not critical in this scenario, those random Verifiers do not have to be located at the edge and can be computationally weak devices. We expect that Contestation is not utilized as its existence ensures that any honest party cannot be falsely convicted of cheating due to an unfortunate initial assignment to dishonest participants.

\section{Threat Model}

In our verification scheme, the Outsourcer, the Contractor, and the Verifier are untrusted and may behave dishonestly. However, we assume that they try to maximize their expected payoff. This type of threat model was first introduced by \cite{belenkiy2008incentivizing}. However verification schemes that assume payoff-maximizing adversaries such as \cite{kupcu} only address a subset of possible threats that can result from these assumptions.   

Due to the lack of a existing collection of threats in this setting, we compile a comprehensive list of possible threats. We consider internal threats of dishonest behavior by one participant and by collusion. Also, we consider threats of external attackers and Quality of Service (QoS) violations such as timeouts and low response rates. Our verification scheme requires all participants to redeem payment, or convicting other parties of cheating by contacting the TTP.

We compiled 9 threats in total. These are described in the following paragraphs. Our verification scheme resists all identified threats with high probability. Table \ref{tab:VerificationSchemeProtocolViolations} summarizes the techniques used by our verification scheme to prevent or detect each possible protocol violation we identified. The confidence column indicates the probability at which the protocol violation can be prevented or detected by our techniques. Note that Contestation provides a 100\% detection rate of associated protocol violations only if more than 50\% of available Verifiers in the ecosystem are non-colluding by not agreeing on an identical incorrect response.

\textbf{Contractor sends back false responses to save resources}

If a Contractor sends back incorrect responses, the Outsourcer detects this with high probability by sending random samples to the Verifier and comparing if responses belonging to the same input from both participants are equal. In the section on sampling, we explained that even with a low number of samples, a cheating Contractor can be detected with nearly 100\% confidence. 

\textbf{Verifier sends back false responses to save resources}

When the Outsourcer detects two unequal responses from the Verifier and the Contractor, our verification scheme provisionally accuses the Contractor of cheating. However, the Contractor can perform our developed Contestation protocol to prove that the Verifier sent the incorrect response instead.  

\textbf{Outsourcer sends back different inputs to Contractor and Verifier to refuse payment}

A cunning dishonest behavior of the Outsourcer that is not addressed by current academic literature is to send two different inputs to the Contractor and the Verifier using the same index. Even if the Contractor and Verifier respond honestly, this almost inevitably leads to unequal responses. This behavior can only be detected if a proof is available that the responses resulted from two different inputs. Thus, not only do the Contractor and Verifier have to sign their responses in our verification scheme, but also the Outsourcer has to sign each sent input with contract-related information. If the Contractor and Verifier also sign the Outsourcer's input signature with their responses, they committed to a specific input rather than only an input index. Thus, it can be verified whether the input signature that the Outsourcer and Verifier committed on was computed based on the same raw input. This verification is performed by the payment settlement entity at the beginning of Contestation.  

\textbf{Contractor or Verifier tries to avoid global penalties when convicted of cheating or QoS violations}

Even when a Verifier or Contractor is detected cheating by an Outsourcer, they may claim never to have sent the reported response. The use of digital signatures prevents this threat. Any behavior aimed to change the record or resubmit a response is detected with 100\% confidence.

\textbf{Participant refuses to pay even if obliged to by the protocol}

Even if the Outsourcer is obliged to reward an honest Contractor or Verifier, there needs to be a way to enforce the payment. Likewise, the Verifier or the Contractor might try to reject paying a penalty fee when detected cheating. 

Microtransactions sent for each response are not an option. Usually, the payment scheme can become a latency bottleneck, and each transaction comes with transaction costs. Therefore, payment has to be handled after a contract ends. Thus, we assume that the payment scheme used along with our verification scheme supports deposits and payment on other participants' behalf. Also, it needs to be able to verify the signatures of the participants. However, it does not have to be able to recompute any values or execute contract-specific functions. Any TTP or Blockchain that supports these requirements can be used with our verification scheme.

\textbf{Outsourcer and Verifier collude to refuse payment and save resources}

The Outsourcer and the Verifier may collude to report the Contractor for cheating. Contestation detects this dishonest behavior. Nevertheless, we provide additional measurements to avoid this type of collusion. With the use of Randomization, the Outsourcer and the Contractor commit on a random Verifier. If the Outsourcer ignores the Randomization protocol and picks a Verifier itself, it is missing the commitment signature of the Contractor, which is revealed in Contestation in case of a dispute. 

Additionally, our verification scheme supports contracts incentives where acting honestly is the dominant strategy. If the Verifier is detected cheating by the Contractor through Contestation, it has to pay a fine. If the incentives are set correctly, acting honestly maximizes the expected payoff. 

\textbf{Contractor and Verifier collude to save resources}

The Contractor and the Verifier may collude to save computational resources by generating an incorrect response and sending it to the Outsourcer. The Outsourcer checks if both results match and assumes the responses to be correct. Our verification scheme prevents the communication of the Verifier and the Contractor through Randomization. 

Additionally, a contract with the right incentives that lead to being honest being a dominant strategy includes a bounty for the Contractor if it detects a cheating Verifier and a bounty for the Verifier to detect a dishonest Contractor. If bounty and fine are set correctly, colluding is not rational for both participants from an individual point of view. This measurement makes ad-hoc collusion between Contractor and Verifier highly unlikely as well.

Beyond our verification scheme, an Outsourcer may decide to utilize more than one Verifier, if possible re-executing inputs by itself at a lower sampling rate, or contacting a TTP after the contracts ends as it still holds all commitments to accuse the other participants of cheating.

\textbf{Timeouts, Low Response Rate, High Response time}

In our verification scheme, dishonest behavior of the Contractor or the Verifier is punished with the refusal of payment, and, if specified in the contract, with a fine. In contrast, QoS violations such as timeouts, low response rate, or high response time come without monetary consequences. We include QoS violations in our threat model, as participants' hardware might be responsible for bad QoS. Therefore, participating in an ecosystem with insufficient processing or networking capabilities should be discouraged. 

We use blacklists, contract abortion, and reviews to punish bad QoS or to promote a good QoS. Whenever a participant receives a message from another participant that exceeds the QoS thresholds specified in its internal parameters, it may abort the current contract due to a QoS violation. It may also blacklist the other participant and submit a negative review on this participant. Thus, a participant who can not provide good QoS misses out on the current contract's ongoing payments and may receive fewer assignments or less reward in the future due to a bad review and blacklisting. Even though a bad QoS might not be the fault of the other participant, frequent negative and positive reviews ensure that the resulting review balance accurately represents the participant's reliableness over time.

\textbf{Message Tampering}

An external attacker may attempt to tamper with messages sent between participants to harm a participant. In our verification scheme, each participant generates a digital signature over each message they send. Thus, any participant receiving a message tampered with by an attacker immediately detects that the message is ill-formatted because the attacker cannot generate a valid signature associated with one of the participants. We acknowledge that by using Merkle Trees for improved efficiency, this leaves room for a message tampering attack that is detected with a slight delay when the proof-of-membership challenge gets sent.

\textbf{Other Threats}

Outsourcers and Contractors can join the ecosystem with an arbitrary amount of identities without hurting the system's security. However, a trusted entity should issue identity checks of new Verifiers to prevent Sybil attacks \cite{nabi}. Otherwise a malicious adversary can increase its probability of matching with colluding participants. The payment settlement entity is ideal for this task as existing payment settlement entities often already provide identity checks.

\begin{table}[]
  \centering
  \caption{Utilized Techniques to Prevent Protocol Violations}
  \resizebox{\textwidth}{!}{
    \begin{tabular}{|p{7em}|p{22em}|p{18.5em}|p{4.5em}|}
    \toprule
    Type of Violation & Description & Techniques & Confidence \\
    \midrule
    Dishonest Behavior by Individual & 1. Contractor sends back false response to save resources & Sampling-based re-execution, utilization of a third party Verifier if required & Up to 100\%  \\
\cmidrule{2-4}   & 2. Verifier sends back false response to save resources & Contestation & 100\% \\
\cmidrule{2-4}     & 3. Outsourcer sends different input to Contractor and Verifier to refuse payment & Digital Signatures (signature chain), Contestation & 100\% \\
\cmidrule{2-4}    & 4. Contractor or Verifier tries to avoid global penalties & Digital Signatures & 100\% \\
\cmidrule{2-4}     & 5. Participant refuses to pay even if obliged to by the protocol & TTP or Blockchain that is authorized to conduct payment on behalf of another entity & 100\% \\
    \midrule
    Dishonest Behavior via Collusion & 6. Outsourcer and Verifier collude to refuse payment and save resources & Randomization, Game-theoretic incentives, Contestation & 100\% \\
\cmidrule{2-4}    & 7. Contractor and Verifier collude to save resources & Randomization, Game-theoretic incentives & High confidence \\
    \midrule
    QoS Violation & 8. Timeout, Low Response Rate, High Response Time & Blacklisting, Review system,  Contract abortion & 100\% \\

    \midrule
    External Threat & 9. Message Tampering & Digital Signatures & 100\% \\
    \bottomrule
    \end{tabular}%
    }
  \label{tab:VerificationSchemeProtocolViolations}%
\end{table}%

\newpage

\section{Related Work}\label{chapter:literature_review}

Related work has identified different components that an edge computing marketplace should provide. These components include a matching and price-finding algorithm \cite{zavodovski}, a payment scheme \cite{rahmani} \cite{zhao2019veriml}, in some cases privacy preservation \cite{zhang} \cite{wangPrivacy} \cite{gheisari}, and a verification scheme \cite{8123913} \cite{huang}. In this work we focus on designing a verification scheme and assume that the other components are present.

Compared to schemes based on cryptographic techniques such as Zero-knowledge proofs, Secure Multiparty Computation, or Fully Homomorphic Encryption, re-execution adds only a negligible computational and network overhead on the task. Re-execution can be implemented in several ways. \cite{di-pietro} propose outsourcing computation to multiple Contractors in a multi-round approach. Different inputs are sent to each Contractor in every round and compared by a trusted master node in case the same input got sent to more than one node. The disadvantage of this scheme is that it requires a large amount of available Contractors in proximity to the Outsourcer.

\cite{kupcu} propose a scheme based on complete re-execution. This scheme's general procedure consists of outsourcing the same job to $n=2$ Contractors and accepting the result if the responses match. If responses do not match, the job gets outsourced again to $n_{new} = n^2$ Contractors until no conflicts arise. This approach is similar yet less efficient than our Contestation protocol. As their scheme does not distinguish between Contractors and Verifiers, the resulting overhead is higher. Also, their scheme relies on a trusted time-stamping server that monitors communication between edge devices. This TTP can end up being the bottleneck of the ecosystem.

\cite{eisele} propose a Smart Contract based protocol running on a Blockchain to set incentives for Outsourcers and Contractors to participate in an outsourcing market place. For verification, their scheme uses semi-trusted third parties. The incentives in the system are set in a way that dishonest behavior is discouraged. For verification, their scheme uses sampling-based re-execution. This scheme's main disadvantage is that current transaction fees in Ethereum are too costly for short contracts. Also, Verifiers have to be partially-trusted.

We conclude that out of these different re-execution approaches, sampling based re-execution is the most practical. We further borrow the following other techniques from related work to improve the outsourcing process.
\begin{enumerate}
    \item Digitial dignatures to authenticate received messages \cite{huang}  \cite{chen} \cite{wei} \cite{nabi}.
    \item Merkle Trees to reduce the required number of digital signatures that need to be sent \cite{huang}  \cite{du} \cite{wei} \cite{nabi}.
    \item TTPs or Blockchains to resolve payments or record communication \cite{huang} \cite{eisele} \cite{chen} \cite{nabi}  \cite{kupcu}.
    \item A global reputation system to promote honest behavior \cite{di-pietro}.
\end{enumerate}

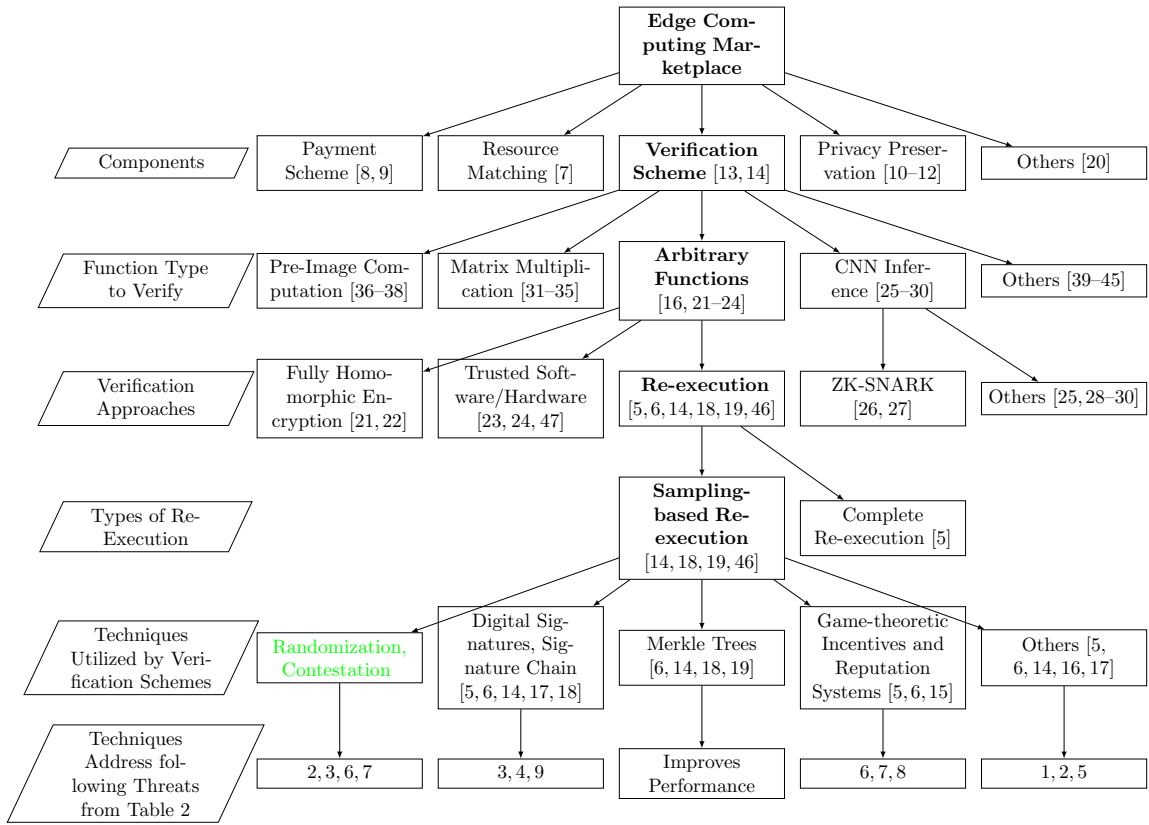
\begin{figure}[!htbp]
    \centering

\begin{adjustbox}{width=\textwidth} 
 
\begin{tikzpicture}[font=\normalsize,scale = 1, every node/.style={scale=1, text width = 3cm, align=center}]
 
\node[draw,
    minimum width=1 cm,
    minimum height=0.5cm] (blockEdge) { \textbf{Edge Computing Marketplace}};

\node[draw,
    below = 1cm of blockEdge,
    minimum width=1 cm,
    minimum height=0.5cm] (blockVer) { \textbf{Verification Scheme} \cite{8123913, huang}};
    
\node[draw,
    left = 0.3cm of blockVer,
    minimum width=1 cm,
    minimum height=0.5cm] (blockRe) { Resource Matching \cite{zavodovski}};

\node[draw,
    left = 0.3cm of blockRe,
    minimum width=1 cm,
    minimum height=0.5cm](blockPay) { Payment Scheme \cite{rahmani, zhao2019veriml}};

\node[draw,
    right=0.3cm of blockVer,
    minimum width=1 cm,
    minimum height=0.5cm] (blockPriv) { Privacy Preservation \cite{zhang, wangPrivacy, gheisari}};
    
    \node[draw,
    right=0.3cm of blockPriv,
    minimum width=1 cm,
    minimum height=0.5cm] (blockOthers) { Others \cite{psaras}};

  
\node[draw,
    below = 1cm of blockVer,
    minimum width=1 cm,
    minimum height=0.5cm] (blockAr) { \textbf{Arbitrary Functions} \cite{arbitrary1, arbitrary2, docker, sgx, eisele} };

\node[draw,
    right = 0.3cm of blockAr,
    minimum width=1 cm,
    minimum height=0.5cm] (blockCNN) { CNN Inference  \cite{neural-network1, neural-network2, neural-network3, neural-network4, neural-network5, neural-network6}};

\node[draw,
    left = 0.3cm of blockAr,
    minimum width=1 cm,
    minimum height=0.5cm] (blockMatrix) { Matrix Multiplication \cite{freivald, matrix-multiplication1, matrix-multiplication2, matrix-multiplication3, matrix-multiplication4} };
    
\node[draw,
    left = 0.3cm of blockMatrix,
    minimum width=1 cm,
    minimum height=0.5cm] (blockPre) { Pre-Image Computation  \cite{Pre-image1, Pre-image2, Pre-image3}};

\node[draw,
    right = 0.3cm of blockCNN,
    minimum width=1 cm,
    minimum height=0.5cm] (blockOtherOps) { Others \cite{other-functions1, other-functions2, linear-equations, matrix-inversion,polynomials1, polynomials2,regression-analysis}};
 

\node[draw,
    below = 1cm of blockAr,
    minimum width=1 cm,
    minimum height=0.5cm] (blockRex) {\textbf{ Re-execution} \cite{huang, du, wei,wang,nabi,kupcu} };
    
    \node[draw,
    left = 0.3cm of blockRex,
    minimum width=1 cm,
    minimum height=0.5cm] (blockTrusted) { Trusted Software/Hardware \cite{docker,sgx_explained, sgx}};
    
\node[draw,
    left = 0.3cm of blockTrusted,
    minimum width=1 cm,
    minimum height=0.5cm] (blockFHE) { Fully Homomorphic
Encryption \cite{arbitrary1,arbitrary2} };
    
\node[draw,
    right = 0.3cm of blockRex,
    minimum width=1 cm,
    minimum height=0.5cm] (blockZK) { ZK-SNARK \cite{neural-network2, neural-network3} };

\node[draw,
    right = 0.3cm of blockZK,
    minimum width=1 cm,
    minimum height=0.5cm] (blockOtherVer) { Others \cite{neural-network1, neural-network4, neural-network5,neural-network6} };


\node[draw,
    below = 1cm of blockRex,
    minimum width=1 cm,
    minimum height=0.5cm] (blockSRex) {\textbf{Sampling-based Re-execution} \cite{huang, du, wei,wang} };
    
\node[draw,
    right = 0.3cm of blockSRex,
    minimum width=1 cm,
    minimum height=0.5cm] (blockCRex) {Complete Re-execution \cite{kupcu}};


\node[draw,
    below = 1cm of blockSRex,
    minimum width=1 cm,
    minimum height=0.5cm] (blockMerkle) {Merkle Trees \cite{huang,du,wei,nabi}};
    
\node[draw,
    right = 0.3cm of blockMerkle,
    minimum width=1 cm,
    minimum height=0.5cm] (blockGame) {Game-theoretic Incentives and Reputation Systems \cite{nabi,di-pietro,kupcu}};

    \node[draw,
    left = 0.3cm of blockMerkle,
    minimum width=1 cm,
    minimum height=0.5cm] (blockSig) {Digital Signatures, Signature Chain \cite{huang, chen, wei, nabi,kupcu} };

    
    \node[draw,
    right = 0.3cm of blockGame,
    minimum width=1 cm,
    minimum height=0.5cm] (blockOtherTech) {Others \cite{huang,eisele,chen, nabi, kupcu}};

\node[draw,
    left = 0.3cm of blockSig,
    minimum width=1 cm,
    minimum height=0.5cm] (blockOur) {\textcolor{green}{ Randomization, Contestation }};
    

\node[draw,
    below = 1.5cm of blockOur,
    minimum width=1 cm,
    minimum height=0.5cm] (blockOurN) { $2,3,6,7$ };
    
\node[draw,
    right = 0.3cm of blockOurN,
    minimum width=1 cm,
    minimum height=0.5cm] (blockSigN) { $3,4,9$ };

\node[draw,
    right = 0.3cm of blockSigN,
    minimum width=1 cm,
    minimum height=0.5cm] (blockMerkleN) { Improves Performance };
    
\node[draw,
    right = 0.3cm of blockMerkleN,
    minimum width=1 cm,
    minimum height=0.5cm] (blockGameN) { $6,7,8$ };


\node[draw,
    right = 0.3cm of blockGameN,
    minimum width=1 cm,
    minimum height=0.5cm] (blockOtherTechN) { $1,2,5$ };




\draw[-latex] (blockEdge) -- (blockPay);
\draw[-latex] (blockEdge) -- (blockVer);
\draw[-latex] (blockEdge) -- (blockRe);
\draw[-latex] (blockEdge) -- (blockPriv);
\draw[-latex] (blockEdge) -- (blockOthers);

\draw[-latex] (blockVer) -- (blockAr);
\draw[-latex] (blockVer) -- (blockCNN);
\draw[-latex] (blockVer) -- (blockPre);
\draw[-latex] (blockVer) -- (blockMatrix);
\draw[-latex] (blockVer) -- (blockOtherOps);

\draw[-latex] (blockAr) -- (blockRex);
\draw[-latex] (blockAr) -- (blockFHE);
\draw[-latex] (blockCNN) -- (blockZK);
\draw[-latex] (blockAr) -- (blockTrusted);
\draw[-latex] (blockCNN) -- (blockOtherVer);

\draw[-latex] (blockRex) -- (blockSRex);
\draw[-latex] (blockRex) -- (blockCRex);

\draw[-latex] (blockSRex) -- (blockOur);
\draw[-latex] (blockSRex) -- (blockSig);

\draw[-latex] (blockSRex) -- (blockMerkle);
\draw[-latex] (blockSRex) -- (blockGame);
\draw[-latex] (blockSRex) -- (blockOtherTech);

\draw[-latex] (blockOur) -- (blockOurN);
\draw[-latex] (blockSig) -- (blockSigN);
\draw[-latex] (blockMerkle) -- (blockMerkleN);
\draw[-latex] (blockGame) -- (blockGameN);
\draw[-latex] (blockOtherTech) -- (blockOtherTechN);

\node[draw,
    trapezium, 
    trapezium left angle = 65,
    trapezium right angle = 115,
    trapezium stretches,
    left = 0.3cm of blockPay,
    minimum width=1 cm,
    minimum height=0.5cm] (blockComp) { Components };

\node[draw,
    trapezium, 
    trapezium left angle = 65,
    trapezium right angle = 115,
    trapezium stretches,
    left = 0.3cm of blockPre,
    minimum width=1 cm,
    minimum height=0.5cm] (blockFun) { Function Type to Verify };
    
\node[draw,
    trapezium, 
    trapezium left angle = 65,
    trapezium right angle = 115,
    trapezium stretches, 
    left = 0.3cm of blockFHE,
    minimum width=1 cm,
    minimum height=0.5cm] (blockApp) { Verification Approaches };
    
\node[draw,
    trapezium, 
    trapezium left angle = 65,
    trapezium right angle = 115,
    trapezium stretches, 
    below = 1.5cm of blockApp,
    minimum width=1 cm,
    minimum height=0.5cm] (blockTypes) { Types of Re-Execution };
    
\node[draw,
    trapezium, 
    trapezium left angle = 65,
    trapezium right angle = 115,
    trapezium stretches,
    left = 0.3cm of blockOur,
    minimum width=1 cm,
    minimum height=0.5cm] (blockTechs) { Techniques Utilized by Verification Schemes };
    
\node[draw,
    trapezium, 
    trapezium left angle = 65,
    trapezium right angle = 115,
    trapezium stretches,
    left = 0.3cm of blockOurN,
    minimum width=1 cm,
    minimum height=0.5cm] (blockTechs) { Techniques Address following Threats from Table \ref{tab:VerificationSchemeProtocolViolations} };

\end{tikzpicture}

\end{adjustbox}
\caption{Overview: Designed verification scheme}
\label{fig:OverviewFlow}
\end{figure}

However, even by combining these existing techniques from different schemes we are only able to solve 7 out of our 9 identified possible threats. We address the remaining two ones by our Randomization and Contestation protocols (marked green in figure \ref{fig:OverviewFlow}). Figure \ref{fig:OverviewFlow} shows an overview of our scheme in the context of our literature analysis. 



\section{Performance}\label{chapter:evaluation}

In our test setup, a Raspberry Pi (Outsourcer) sends a real-time webcam stream to two different machines (Verifier and Contractor) in the local network. The Contractor and the Verifier send back bounding boxes of detected objects in the frames.
One test setup uses a regular GPU/CPU for inference, and one uses a Coral USB Accelerator. The Coral USB Accelerator is an entry-level Tensor Processing Unit (TPU) that is specifically designed to perform neural network inference \cite{CoralUSB}. We used weights that were trained on the Microsoft Coco dataset \cite{Coco}. We used Yolov4 \cite{YOLOv4} and MobileNet SSD V2 \cite{ssd} as models for object detection.

Our implementation uses the NaCl ED25519 signature scheme and can perform the verification scheme's task in parallel to the object detection. This eliminates the latency overhead of our scheme as the verification scheme's task utilize one CPU thread while the GPU is the bottleneck when performing a single object detection. Table \ref{tab:KeyResults} shows the key results of our test implementation. The source code of our implementation is publicly available \href{https://github.com/chart21/Verification-of-Outsourced-Object-Detection}{here}. Our implementation supports multithreading, Merkle Trees, and non-blocking message pattern to improve the efficiency of our scheme. A diagram of the test setup is provided in the appendix.

\begin{table*}[!h]

  \centering
  \caption{Key Results}
  \resizebox{\textwidth}{!}{
    \begin{tabular}{|p{4.835em}|p{5.75em}|p{3.25em}|p{5em}|p{6em}|p{3em}|p{6em}|p{5.7em}|p{5.7em}|}
    \toprule
    Participant & Device & CPU & GPU & Model & Frames per second & Time spent on application processing (\%) & Time spent on verification scheme (\%) & Time spent on verification scheme (ms) \\
    \midrule
    Outsourcer & Raspberry Pi \newline{}Model 4B &  &  & MobileNet SSD V2 300$\times$300 & 236.00 & 78.70 & 21.30 & 0.90 \\
    \midrule
    Outsourcer & Raspberry Pi \newline{}Model 4B &  &  & Yolov4 tiny\newline{}416$\times$416 & 146.90 & 85.10 & 14.90 & 1.01 \\
    \midrule
     
    Contractor & Desktop PC & Core i7 3770K & GTX 970 & Yolov4 tiny\newline{}416$\times$416 & 68.06 & 100.00 & 0.00 & 0.00 \\
     \midrule
    Contractor & Desktop PC & Core i7 3770K & Coral USB Accelerator & MobileNet SSD V2 300$\times$300 & 63.59 & 100.00 & 0.00 & 0.00 \\
     \midrule
    Contractor & Notebook & Core i5 4300U & Coral USB Accelerator & MobileNet SSD V2 300$\times$300 & 49.30 & 100.00 & 0.00 & 0.00 \\
     \midrule
    Verifier & Notebook & Core i5 4300U & Coral USB Accelerator & MobileNet SSD V2 300$\times$300 & 28.75 & -  & 0.00     & 0.00 \\
    \bottomrule
    \end{tabular}%
    }
  \label{tab:KeyResults}%

\end{table*}

Our results show that our verification scheme causes less than 1ms of latency overhead per frame. The Contractor's GPU is the system's bottleneck in our test setup, limiting overall performance to 68.06 fps. As we only used a mid-range GPU and an entry-level edge accelerator in our tests, more potent Contractor hardware could increase overall system performance to match the Outsourcer performance of more than 200fps. The average 416x416 frame has a size of 120 KB in our tests. The Network bandwidth overhead per participant is negligible at a maximum of 84 bytes per frame. It consists of a 512-bit large signature, and at most, five 32 bit integers such as frame index, acknowledged responses, and other contract-related information. When Merkle Trees are utilized, the Contractor and the Verifier only send signatures when the Outsourcer requests a proof-of-membership challenge.


\section{Conclusion}\label{chapter:Conclusion}

In this work, we proposed a scheme for verifying outsourced arbitrary functions in a computation marketplace. The verification scheme is resistant to a comprehensive list of protocol violations that might occur in a computation marketplace with untrusted participants.
We benchmarked our verification scheme's performance on consumer hardware and TPUs. Our verification scheme achieves less than 1ms of latency overhead per frame on all tested machines. By utilizing concurrency, the overhead can be reduced to 0 by running the verification scheme's tasks in a parallel thread to the outsourced computation. The network bandwidth overhead of our scheme caused mainly by digital signatures is negligible (at most 84 bytes per frame).

In comparison with verification schemes proposed by the current academic literature, our verification scheme provides additional security by preventing or detecting all our identified protocol violations. At the same time it only requires third-party involvement outside the edge. For ensuring high performance, our implementation supports non-blocking message patterns, Merkle trees, multi-threading, and the most efficient digital signature library in Python according to our benchmarks (NaCl ED25519). These performance and security characteristics make our scheme an ideal choice to be used within a latency-sensitive edge computing marketplace that matches untrusted third-party resources to computationally weak, untrusted IoT devices. 

Our verification scheme implements one essential component of a fully functioning edge computing marketplace. As illustrated in figure \ref{fig:OverviewFlow}, the remaining components to make an edge computing marketplace viable are: Payment, Matching and price-finding, and Privacy preservation. Future work may first identify the best solution proposed for each component and then aggregate these components to an end-to-end system that can be deployed as a standalone edge computing marketplace for arbitrary functions.

\bibliographystyle{IEEEtran}
\bibliography{refs}
\section{Appendix}

Codebase: \url{https://github.com/chart21/Verification-of-Outsourced-Object-Detection}

\begin{figure*}[ht]
	\centering
  \includegraphics[angle=90, width=0.65\textwidth]{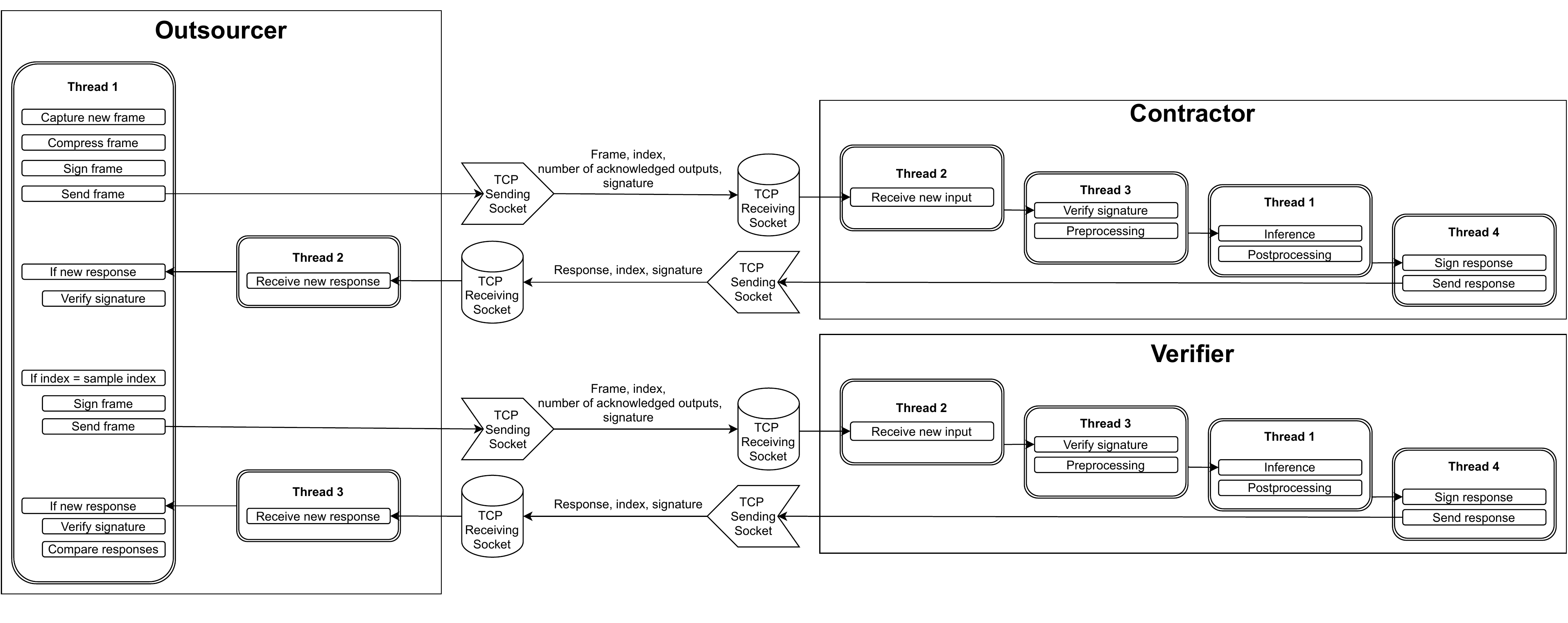}
	\caption{Test Setup}
	\label{multi-threadingCPUGPU}
\end{figure*}

\end{document}